\documentclass[MSNbibl,nameyear,dvips]{arxstspdf}
\usepackage{flushend}
\usepackage{stfloats}

\volume{27}
\issue{1}
\pubyear{2012}
\firstpage{1}
\lastpage{2}
\doi{10.1214/11-STS385}

\begin{document}
\begin{frontmatter}
\vspace*{6pt}
\title{A Tribute to Charles Stein}
\runtitle{A Tribute to Charles Stein}

\begin{aug}
\author[a]{\fnms{Edward I.} \snm{George}\corref{}\ead[label=e1]{edgeorge@wharton.upenn.edu}}
\and
\author[b]{\fnms{William E.} \snm{Strawderman}\ead[label=e2]{straw@stat.rutgers.edu}}
\runauthor{E. I. George and W. E. Strawderman}

\affiliation{The Wharton School and Rutgers University}

\address[a]{Edward I. George is Professor, Department of Statistics,
The Wharton School, Philadelphia, Pennsylvania 19104-6340, USA \printead{e1}.}
\address[b]{William E. Strawderman is Professor, Department of
Statistics, Rutgers University,
Piscataway, New Jersey 08854-8019, USA \printead{e2}.}

\end{aug}



\end{frontmatter}

In 1956, Charles Stein published an article that was to forever change
the statistical approach to high-dimensional estimation. His stunning
discovery that the usual estimator of the normal mean vector could be
dominated in dimensions 3 and higher amazed many at the time, and
became the catalyst for a vast and rich literature of substantial
importance to statistical theory and practice. As a~tribute to Charles
Stein, this special issue on minimax shrinkage estimation is devoted to
developments that ultimately arose from Stein's investigations into
improving on the UMVUE of a multivariate normal mean vector. Of course,
much of the early literature on the subject was due to Stein himself,
including a~key technical lemma commonly referred to as Stein's Lemma,
which leads to an unbiased estimator of the risk of an almost arbitrary
estimator of the mean vector.

The following ten papers assembled in this volume represent some of the
many areas into which shrinkage has expanded (a one-dimensional pun, no
doubt). Clearly, the shrinkage literature has bran\-ched out
substantially since 1956, the many contributors and the breadth of
theory and practice being now far too large to cover with any degree of
completeness in a review issue such as this one. But what these papers
do show is the lasting impact of Stein (\citeyear{Ste56}), and the ongoing vitality
of the huge area that he catalyzed.

\begin{itemize}

\item Berger, Jefferys and M\"uller (Bayesian nonparametric shrinkage
applied to Cepheid Star oscillations) model Cepheid star oscillations
via Bayesian analysis\vadjust{\goodbreak} of a wavelet expansion. They illustrate how two
types of shrinkage occur, the shrinkage of parameters toward some
prespecified subsets, and the setting of some parameters to zero, which
in this setting induces smoothness of the function estimate.

\item Brandwein and Strawderman (Stein estimation for spherically
symmetric distributions: Recent developments) update an earlier
\textit{Statistical Science} review paper (Brandwein and Strawderman, \citeyear{BraStr90}).
Going further, this paper emphasizes the distributional robustness
properties of a class of Stein estimators when a residual vector is
available to estimate scale.

\item Brown and Zhao (A geometrical explanation of Stein shrinkage)
flesh out Stein's initial geometric heuristic for the inadmissibility
of the usual estimator. By exploiting the spherical symmetry of the
problem and reducing it conceptually to a~two-dimensional framework,
the geometry is cla\-rified to provide justification for the fact that
dimension 3 is the critical threshold for inadmissibility.

\item Cai (Minimax and adaptive inference in nonparametric function
estimation) takes us into the realm of nonparametric function
estimation where minimax shrinkage estimation has turned out to play a
major role. Cai describes how
such function estimation problems are equivalent to problems of
estimating infinite-dimensional multivariate normal means constrained
to lie within compact subsets, and how three different but connected
problems lead to similar minimax results but strikingly different
adaptivity results.

\item Casella and Hwang (Shrinkage confidence procedures) review the
rich developments in confidence set estimation that have paralleled
developments in shrinkage point estimation. Possible improvements over
the classical equivariant region include not only recentering the usual
confidence sets at shrinkage estimators, but also shrinking the volu\-me
of the set, and/or changing the shape of the~set.

\item Fourdrinier and Wells (On loss estimation) discuss the issue of
postdictive assessment of statistical procedures through the study of
loss estimation. One key connection in this literature which links it
with much of classical shrinkage literature is the use of differential
identities and inequalities related to the standard Stein identities,
but here the expressions are typically of higher order. You will
especially love the bi-Laplacian!

\item George, Liang and Xu (From minimax shrinkage estimation to
minimax shrinkage prediction) describe the parallels between the
developments of minimax mean estimation under quadratic risk and
minimax predictive density estimation under Kullback--Leibler risk. The
strong connections reveal a remarkable similarity in the development of
risk calculations between the two problems that allows, at least in
some cases, an almost immediate proof of domination of a Bayesian
shrinkage predictive density estimator over the usual best equivariant one.

\item Ghosh and Datta (Small area shrinkage estimation) study the
practice of ``borrowing strength'' as it relates to small area
estimation, where survey data gathered for decision making at, say, the
state level must be also used to make decisions at the county or town
level. As the amount of survey data at the town level is often quite
small, estimates based only on local data tend to be overly variable.
The use of shrinkage techniques described by the authors is a practical
tool of great utility in this setting.

\item Morris and Lysy (Shrinkage estimation in multilevel normal
models) lay out the evolution of the hierarchical Bayes approach to
shrinkage estimation which was catalyzed by Stein's early work. Moving
through the equal variance to the unequal variance case, through
multilevel model formulations,\ they describe and illustrate the rich
variety of shrinkage estimators and tools that have evolved for their
evaluation.

\item Perlman and Chaudhuri (Reversing the Stein effect) provide
further insight into one of the most shocking aspects of Stein's
initial discoveries, that one could obtain risk improvement over
equivariant estimators near any preselected target, thereby showing
that prior information should not be ignored. They offer a cautionary
tale about what can happen when attempting to relax the requirement
that the shrinkage target selection be independent of the data.

\end{itemize}


%

%

\end{document}